\documentclass{elsart1p}
\usepackage{graphicx}

\usepackage{amssymb}
\begin{document}
%

%
%
\def\d0{D\O}
\def\D0{D\O}
\def\w{$W$}
\def\W{$W$}
\def\z{$Z$}
\def\Z{$Z$}
\def\wg{$W\gamma$}
\def\ww{$WW$}
\def\wpwm{$W^+W^-$}
\def\wz{$WZ$}
\def\wwg{$WW\gamma$}
\def\wwv{$WWV$}
\def\wwz{$WWZ$}
\def\zg{$Z\gamma$}
\def\zzg{$ZZ\gamma$}
\def\zgg{$Z\gamma\gamma$}
\def\pt{$p_T$}
\def\et{$E_T$}
\newcommand{\Eslash}{\mbox{$E \kern-0.6em\slash$}}
\newcommand{\etmis}{\mbox{$\Eslash_T$}}
\def\pdf    {parton distribution function }
\def\pdfs   {parton distribution functions }
\def\ifmath#1{\relax\ifmmode #1\else $#1$}%
\def\TeV{\ifmmode {\mathrm{ Te\kern -0.1em V}}\else
                   \textrm{Te\kern -0.1em V}\fi}%
\def\GeV{\ifmmode {\mathrm{ Ge\kern -0.1em V}}\else
                   \textrm{Ge\kern -0.1em V}\fi}%
\def\MeV{\ifmmode {\mathrm{ Me\kern -0.1em V}}\else
                   \textrm{Me\kern -0.1em V}\fi}%
\def\GeVcc{\ifmmode {\mathrm{ \GeV/c^2}}\else
                   \textrm{Ge\kern -0.1em V/c$^2$}\fi}%
\def\MeVcc{\ifmmode {\mathrm{ \MeV/c^2}}\else
                   \textrm{Me\kern -0.1em V/c$^2$}\fi}%
\def\pbar               {\mbox{$\overline{p}$}}
\def\pbarp              {\mbox{$\overline{p}p$}}
\def\ppbar              {\mbox{$p\overline{p}$}}
\def\ttbar              {\mbox{$t\overline{t}$}}
\newcommand{\Afb}       {A_{\mathrm{FB}}}
\newcommand{\AFB}       {A_{\mathrm{FB}}}
\newcommand{\ee}        {\mbox{$e^+e^-$}}
\newcommand{\qq}        {\mbox{$q\overline q$}}
\newcommand{\uu}        {\mbox{$u\overline u$}}
\newcommand{\dd}        {\mbox{$d\overline d$}}
\newcommand{\bb}        {\mbox{$b\overline b$}}
\newcommand{\SM}        {\mbox{SM}}
\def\eg{{\it e.g.}}
\def\ie{{\it i.e.}}
\def\etal{{\it et al.}}
\def\isajet{{\sc isajet}}
\def\geant{{\sc geant}}
\def\pythia{{\sc pythia}}
\def\vecbos{{\sc vecbos}}
\def\herwig{{\sc herwig}}
\newcommand{\tab}{\hspace*{0.5in}}

\def\grad{\nabla}
\def\gradvec{\mbox{\boldmath $\nabla$}}
\def\Aslash{\mbox{${\hbox{$A$\kern-0.55em\hbox{/}}}$}}
\def\pslash{\mbox{${\hbox{$p$\kern-0.45em\hbox{/}}}$}}


%
%
\def\andit{{\it\&}}
\def\half{\frac{1}{2}}
\def\thalf{\tfrac{1}{2}} 
\def\third{\frac{1}{3}}
\def\quarter{\frac{1}{4}}
\def\rii{\sqrt{2}}
\def\to{\rightarrow}
\def\S{\mathhexbox279}
\def\gesim{\,{\raise-3pt\hbox{$\sim$}}\!\!\!\!\!{\raise2pt\hbox{$>$}}\,}
\def\lesim{\,{\raise-3pt\hbox{$\sim$}}\!\!\!\!\!{\raise2pt\hbox{$<$}}\,}
\def\boldoverdot{\,{\raise6pt\hbox{\bf.}\!\!\!\!\>}}
\def\re{{\bf Re}}
\def\im{{\bf Im}}
\def\ie{{i.e.}}
\def\cf{{\it cf.}\ }
\def\ibid{{\it ibid.}\ }
\def\etal{{\it et. al.}}
\def\acal{{\cal A}}
\def\bcal{{\cal B}}
\def\ccal{{\cal C}}
\def\dcal{{\cal D}}
\def\ecal{{\cal E}}
\def\fcal{{\cal F}}
\def\gcal{{\cal G}}
\def\hcal{{\cal H}}
\def\ical{{\cal I}}
\def\jcal{{\cal J}}
\def\kcal{{\cal K}}
\def\lcal{{\cal L}}
\def\mcal{{\cal M}}
\def\ncal{{\cal N}}
\def\ocal{{\cal O}}
\def\pcal{{\cal P}}
\def\qcal{{\cal Q}}
\def\rcal{{\cal R}}
\def\scal{{\cal S}}
\def\tcal{{\cal T}}
\def\ucal{{\cal U}}
\def\vcal{{\cal V}}
\def\wcal{{\cal W}}
\def\xcal{{\cal X}}
\def\ycal{{\cal Y}}
\def\zcal{{\cal Z}}
\def\iBB{ \hbox{{\mysmallii I}}\!\hbox{{\mysmallii I}} }
\def\bBB{ \hbox{{\mysmallii I}}\!\hbox{{\mysmallii B}} }
\def\pBB{ \hbox{{\mysmallii I}}\!\hbox{{\mysmallii P}} }
\def\rBB{ \hbox{{\mysmallii I}}\!\hbox{{\mysmallii R}} }
\def\alpbf{{\pmb{$\alpha$}}}
\def\betbf{{\pmb{$\beta$}}}
\def\gambf{{\pmb{$\gamma$}}}
\def\delbf{{\pmb{$\delta$}}}
\def\epsbf{{\pmb{$\epsilon$}}}
\def\zetbf{{\pmb{$\zeta$}}}
\def\etabf{{\pmb{$\eta$}}}
\def\thebf{{\pmb{$\theta$}}}
\def\varthebf{{\pmb{$\vartheta$}}}
\def\iotbf{{\pmb{$\iota$}}}
\def\kapbf{{\pmb{$\kappa$}}}
\def\lambf{{\pmb{$\lambda$}}}
\def\mubf{{\pmb{$\mu$}}}
\def\nubf{{\pmb{$\nu$}}}
\def\xibf{{\pmb{$\xi$}}}
\def\pibf{{\pmb{$\pi$}}}
\def\varpibf{{\pmb{$\varpi$}}}
\def\rhobf{{\pmb{$\rho$}}}
\def\sigbf{{\pmb{$\sigma$}}}
\def\taubf{{\pmb{$\tau$}}}
\def\upsbf{{\pmb{$\upsilon$}}}
\def\phibf{{\pmb{$\phi$}}}
\def\varphibf{{\pmb{$\varphi$}}}
\def\chibf{{\pmb{$\chi$}}}
\def\psibf{{\pmb{$\psi$}}}
\def\omebf{{\pmb{$\omega$}}}
\def\Gambf{{\pmb{$\Gamma$}}}
\def\Delbf{{\pmb{$\Delta$}}}
\def\Thebf{{\pmb{$\Theta$}}}
\def\Lambf{{\pmb{$\Lambda$}}}
\def\Xibf{{\pmb{$\Xi$}}}
\def\Pibf{{\pmb{$\Pi$}}}
\def\Sigbf{{\pmb{$\sigma$}}}
\def\Upsbf{{\pmb{$\Upsilon$}}}
\def\Phibf{{\pmb{$\Phi$}}}
\def\Psibf{{\pmb{$\Psi$}}}
\def\Omebf{{\pmb{$\Omega$}}}
\def\ssb{spontaneous symmetry breaking}
\def\vev{vacuum expectation value}
\def\irrep{irreducible representation}
\def\lhs{left hand side\ }
\def\rhs{right hand side\ }
\def\Ssb{Spontaneous symmetry breaking\ }
\def\Vev{Vacuum expectation value}
\def\Irrep{Irreducible representation}
\def\Lhs{Left hand side\ }
\def\Rhs{Right hand side\ }
\def\tr{ \hbox{tr}}
\def\det{\hbox{det}}
\def\Tr{ \hbox{Tr}}
\def\Det{\hbox{Det}}
\def\diag{\hbox{\diag}}
\def\sm{Standard Model}
\def\ev{\hbox{eV}}
\def\kev{\hbox{keV}}
\def\mev{\hbox{MeV}}
\def\gev{\hbox{GeV}}
\def\tev{\hbox{TeV}}
\def\milm{\hbox{mm}}
\def\cm{\hbox{cm}}
\def\m{\hbox{m}}
\def\km{\hbox{km}}
\def\gr{\hbox{gr}}
\def\kg{\hbox{kg}}
%
%
\def\noteeye{ {{$\quad(\!(\subset\!\!\!\!\bullet\!\!\!\!\supset)\!)\quad$}}}
\def\note#1{{\bf \noteeye\nobreak #1 \noteeye } }
\def\doubleundertext#1{
{\undertext{\vphantom{y}#1}}\par\nobreak\vskip-\the\baselineskip\vskip4pt%
\undertext{\hbox to 2in{}}}
\def\inbox#1{\vbox{\hrule\hbox{\vrule\kern5pt
     \vbox{\kern5pt#1\kern5pt}\kern5pt\vrule}\hrule}}
\def\sqr#1#2{{\vcenter{\hrule height.#2pt
      \hbox{\vrule width.#2pt height#1pt \kern#1pt
         \vrule width.#2pt}
      \hrule height.#2pt}}}
\def\today{\ifcase\month\or
  January\or February\or March\or April\or May\or June\or
  July\or August\or September\or October\or November\or December\fi
  \space\number\day, \number\year}
\def\pmb#1{\setbox0=\hbox{#1}%
  \kern-.025em\copy0\kern-\wd0
  \kern.05em\copy0\kern-\wd0
  \kern-.025em\raise.0433em\box0 }
\def\up#1{^{\left( #1 \right) }}
\def\lowti#1{_{{\rm #1 }}}
\def\inv#1{{\frac{1}{#1}}}
\def\deriva#1#2#3{\left({\frac{\partial #1}{\partial #2}}\right)_{#3}}
\def\su#1{{SU(#1)}}
\def\ui{U(1)}
\def\antes{}
\def\despues{.}
\def\dss{ {}^2 }
%
\def\sumprime_#1{\setbox0=\hbox{$\scriptstyle{#1}$}
  \setbox2=\hbox{$\displaystyle{\sum}$}
  \setbox4=\hbox{${}'\mathsurround=0pt$}
  \dimen0=.5\wd0 \advance\dimen0 by-.5\wd2
  \ifdim\dimen0>0pt
  \ifdim\dimen0>\wd4 \kern\wd4 \else\kern\dimen0\fi\fi
\mathop{{\sum}'}_{\kern-\wd4 #1}}
\def\dodraft#1{
\typeout{}
\typeout{**************|||||||||******************}
\typeout{}
\typeout{Paper: #1 (DRAFT)}
\typeout{}
\typeout{**************|||||||||******************}
\typeout{}
\typeout{}
}
%
%
\font\sanser=cmssq8
\font\sanseru=cmssq8 scaled\magstep1 
\font\sanserd=cmssq8 scaled\magstep2 
\font\sanseri=cmssq8 scaled\magstep1
\font\sanserii=cmssq8 scaled\magstep2
\font\sanseriii=cmssq8 scaled\magstep3
\font\sanseriv=cmssq8 scaled\magstep4
\font\sanserv=cmssq8 scaled\magstep5
\font\tyt=cmtt10
\font\tyti=cmtt10 scaled\magstep1
\font\tytii=cmtt10 scaled\magstep2
\font\tytiii=cmtt10 scaled\magstep3
\font\tytiv=cmtt10 scaled\magstep4
\font\tytv=cmtt10 scaled\magstep5
\font\slanti=cmsl10 scaled\magstep1
\font\slantii=cmsl10 scaled\magstep2
\font\slantiii=cmsl10 scaled\magstep3
\font\slantiv=cmsl10 scaled\magstep4
\font\slantv=cmsl10 scaled\magstep5
\font\bigboldi=cmbx10 scaled\magstep1
\font\bigboldii=cmbx10 scaled\magstep2
\font\bigboldiii=cmbx10 scaled\magstep3
\font\bigboldiv=cmbx10 scaled\magstep4
\font\bigboldv=cmbx10 scaled\magstep5
\font\mysmall=cmr8
\font\mysmalli=cmr8 scaled\magstep1
\font\mysmallii=cmr8 scaled\magstep2
\font\mysmalliii=cmr8 scaled\magstep3
\font\mysmalliv=cmr8 scaled\magstep4
\font\mysmallv=cmr8 scaled\magstep5
\font\ital=cmti10
\font\itali=cmti10 scaled\magstep1
\font\italii=cmti10 scaled\magstep2
\font\italiii=cmti10 scaled\magstep3
\font\italiv=cmti10 scaled\magstep4
\font\italv=cmti10 scaled\magstep5
\font\smallit=cmmi7
\font\smalliti=cmmi7 scaled\magstep1
\font\smallitii=cmmi7 scaled\magstep2
\font\rmi=cmr10 scaled\magstep1
\font\rmii=cmr10 scaled\magstep2
\font\rmiii=cmr10 scaled\magstep3
\font\rmiv=cmr10 scaled\magstep4
\font\rmv=cmr10 scaled\magstep5
\font\eightrm=cmr8

\begin{frontmatter}
%
%
%
\title{Measurement of $B_s$ Oscillations and CP Violation Results from \d0 }
%
%
\author{J. Ellison, for the \d0\ Collaboration}
\address{University of California, Riverside, CA 92521, USA}
\begin{abstract}
We present a measurement of the $B_s^0 - \bar B_s^0$ oscillation frequency, $\Delta m_s$, using a combination of semi-leptonic and hadronic  $B_s$ decay candidates selected from data collected by the \d0\ Experiment at the Fermilab Tevatron. We also present several results on CP violation, including an improved measurement of the $B_s$ CP-violating phase from a flavor-tagged analysis of $B_s^0 \to J/\psi + \phi$  decays.
\end{abstract}
\end{frontmatter}
%

\section{Introduction}

Among the primary goals of the Tevatron is the observation of $B_s$ oscillations and the search for CP violation in $B$-meson decays. In this paper we describe some recent results in these areas. The data used for the results described here are based on approximately $1.2 - 2.8$~fb$^{-1}$ of data collected by the \d0\ Experiment at the Fermilab Tevatron. The \d0\ detector consists of a central tracking system surrounded by a uranium liquid-argon calorimeter and an outer muon detection system and is described in~\cite{d0det}.

\section{Measurement of $B_s$ Oscillations}
\label{sec:bosc}

In this paper we report a preliminary measurement of the $B_s^0 - {\bar B}_s^0$ oscillation frequency $\Delta m_s$ based on the analysis of five decay channels:
(1) $B_s^0 \to D_s^- \mu^+ \nu X, D_s^- \to \phi \pi^-$;
(2) $B_s^0 \to D_s^- e+ \nu X, D_s^- \to \phi \pi^-$; 
(3) $B_s^0 \to D_s^- \mu^+ \nu X, D_s^- \to K^{*0} K^-$;
(4) $B_s^0 \to D_s^- \mu^+ \nu X, D_s^- \to K_s^0 K^-$;
(5) $B_s^0 \to D_s^- \pi^+, D_s \to \phi \pi^-$.
The datasets were approximatley 2.4~fb$^{-1}$ for all channels except (4) whihc used a dataset of
1.2~fb$^{-1}$. \d0\ has previously reported direct limits on $B_s$ oscillations using 1~fb$^{-1}$ of data in the first decay channel~\cite{d0_bsosc}. 

After selection and reconstruction of the $B_s^0$ decays, the flavor of the $B_s^0$ at decay is determined from the charge of the muon, electron, or pion in the final state. The flavor at production is determined using a combined flavor tagger based on opposite-side and same-side tagging~\cite{flavor_tag}. The effectiveness of the tagging algorithm is characterized by the tagging efficiency $\epsilon$ and the dilution $\cal D$ defined by ${\cal D} = (N_{RS} - N_{WR}) / (N_{RS} + N_{WR})$, where $N_{RS}$ and $N_{WR}$ are the number of right-sign and wrong-sign tags, respectively. The overall effectiveness of the combined tagger, obtained using $B \to \mu \nu D^{*-}$ data and $B_s \to J/\psi \phi$ data and Monte Carlo, is $\epsilon {\cal D}^2 = 4.49 \pm 0.88 \%$. 

The determination of the proper decay length takes into account possible missing momentum, e.g. due to the presence of a neutrino in the semileptonic decay modes. The proper decay length is given by 
$c \tau_{B_s^0} = x^M K$, where $x^M$ is the measured or visible proper decay length, given by
\begin{eqnarray*}
  x^M = \left[ \frac{{\vec d}_T^{~B_s^0} \cdot {\vec p}_T^{~\ell D_s^-} }{(p_T^{\ell D_s^-})^2} \right] c M_{B_s^0}
  ~~~~~{\rm with} ~~K = \frac{p_T^{\mu D_s^-}}{p_T^{B_s^0}}
 \end{eqnarray*}
 
\noindent
The $K$-factor, $K$, is determined from Monte Carlo.

The measurement of $\Delta m_s$ is accomplished using an unbinned likelihhod fit to the data. The likelihood function is constructed using the probabilities for mixed and unmixed decays,
\begin{eqnarray*}
  P_{\rm mix} (t) = \half \frac{K}{c \tau_{B_s}} \exp \left( - \frac{Kx}{c \tau_{B_s}} \right)
  (1 - {\cal D} \cos{\Delta m_s Kx/c} )\\
  P_{\rm nomix}(t) = \half \frac{K}{c \tau_{B_s}} \exp \left( - \frac{Kx}{c \tau_{B_s}} \right)
  (1 + {\cal D} \cos{\Delta m_s Kx/c} )
\end{eqnarray*}

\noindent
and accounting for detector resolutions and background contributions. The combined amplitude scan, where the likelihood is modified to include an amplitude term, ${\cal L} \propto 1 \pm {\cal A} {\cal D} \cos(\Delta m_s K x / c)$, is shown in Fig.~\ref{fig:osc_likelihood}(a). The likelihood as a function of $\Delta m_s$ is shown in Fig.~\ref{fig:osc_likelihood}(b). The measured value of $\Delta m_s$ is obtained from a fit to the likelihood vs. $\Delta m_s$ in the region of the minimum shown in Fig.~\ref{fig:osc_likelihood}(b), and yields $\Delta m_s = 18.53 \pm 0.93 {\rm{ (stat.)}} \pm 0.30 {\rm (syst.)}$~ps$^{-1}$. The significance of this result is 2.9~$\sigma$. 
\begin{figure}[htp]
\begin{center}
\begin{tabular}{c c}
	\includegraphics[width=0.5\textwidth]{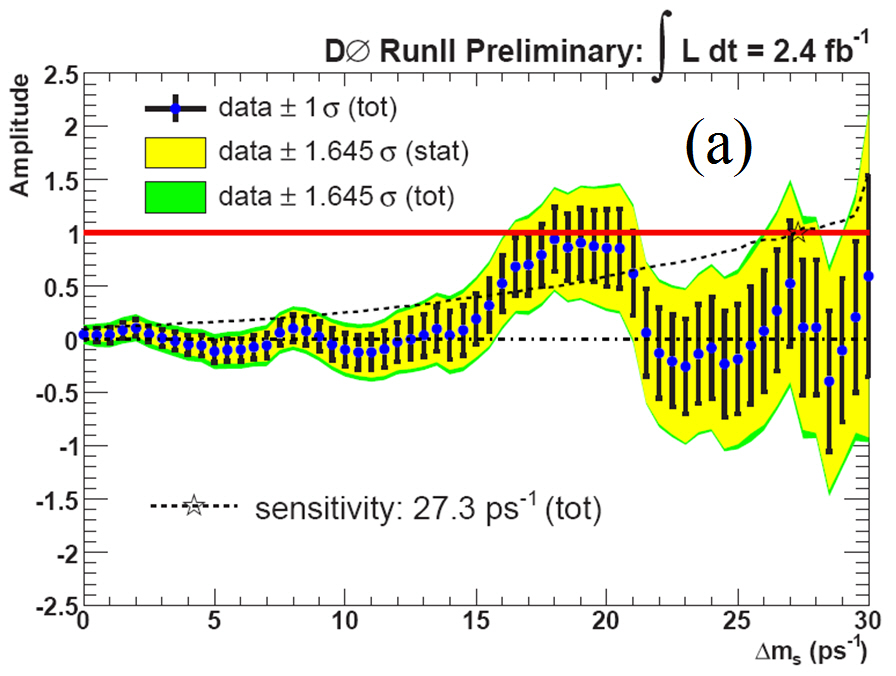} &
	\includegraphics[width=0.5\textwidth]{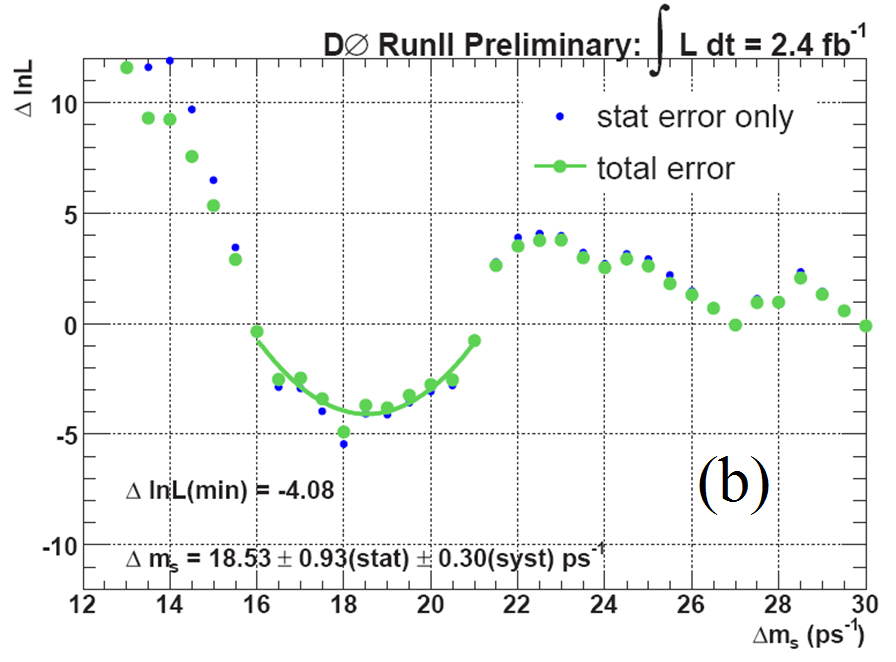}
\end{tabular}
\end{center}
	\caption{\label{fig:osc_likelihood}
	(a) $B_s^0$ oscillation amplitude as a function of $\Delta m_s$ and (b) $B_s^0$ oscillation likelihood  as a function of    $\Delta m_s$.}
\end{figure}

The $\Delta m_s$ measurement can be translated into a constraint on $|V_{td}/V_{ts}|$, and we obtain
\begin{eqnarray*}
\left| \frac{V_{td}}{V_{ts}} \right| = 0.2018 \pm 0.005 \rm{(exp.)} \; ^{+0.0081} _{-0.0060} \rm{(theor.)} 
\end{eqnarray*}

\section{Analysis of $B_d^0 \to J/\psi + K^{*0}$ and $B_s^0 \to J/\psi + \phi$ Decays}

Before discussing CP Violation in $B_s^0 \to J/\psi + \phi$, we present results of an analysis of $B_d^0 \to J/\psi + K^{*0}$ and $B_s^0 \to J/\psi + \phi$ decays, in which CP violation is neglected. The aim is to measure the parameters which describe the time-dependent angular distributions of these decays where the initial $B$ meson flavor is not determined. The relevant parameters are the linear polarization amplitudes, the strong phases, and the lifetimes. This allows us to obtain the lifetime ratio, and test $SU(3)$ flavor symmetry and factorization. This is achieved by fitting to the mass, lifetime, and angular distributions of the decay products. For details of the analysis see~\cite{bs_bd_prl}.

The preliminary results are shown in Table~\ref{tab:table1}. The $B_d$ results are competitive and consistent with previous measurements by CDF, BaBar, and Belle. The measurements of the strong phases indicate the presence of final-state interactions for $B_d^0 \to J/\psi + K^{*0}$, since $\delta_1$ is 3.5~$\sigma$ away from zero. Also, comparison of the polarization amplitudes and strong phases shows no evidence for violation of $SU(3)$ flavor symmetry.
\begin{table}[htp]
	\caption{\label{tab:table1}
	Preliminary results for the linear polarization amplitudes, strong phases, lifetime and width difference for $B_d^0 \to J/\psi + K^{*0}$ and $B_s^0 \to J/\psi + \phi$. $A_0$ and $A_\parallel$ are the linear polarization amplitudes, $\delta_1$, $\delta_2$, and $\delta_\parallel$ are the strong phases, $\tau$ is the lifetime, and for the $B_s$ decay, $\Delta \Gamma_s$ is the width difference between the light and heavy mass eigenstates.}
\begin{center}
	\includegraphics[width=0.9\textwidth]{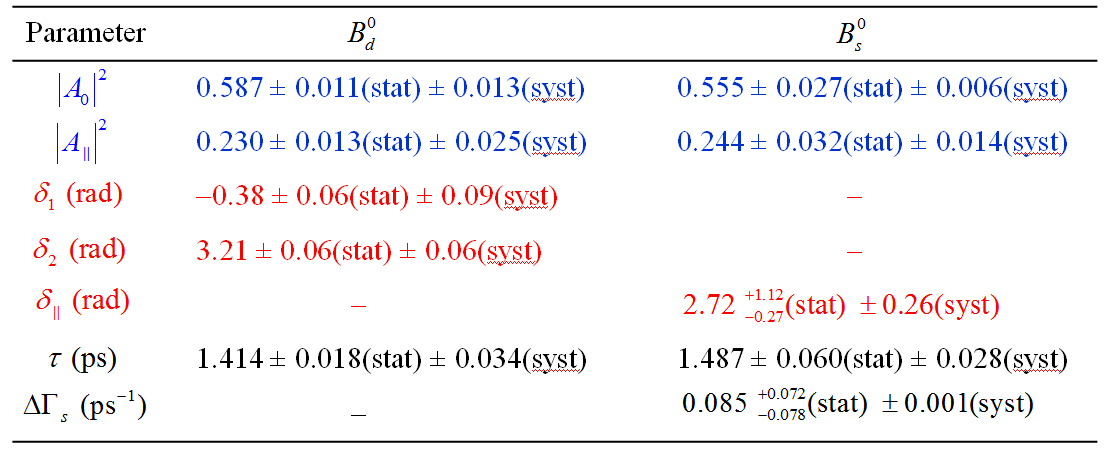}
	\end{center}
\end{table}

\section{Flavor-Tagged Analysis of $B_s^0 \to J/\psi + \phi$ Decays}

The $B_s^0 \to J/\psi \; \phi$ decay involves a final state that is a mixture of CP-even and CP-odd states. In order to separate the CP-even and CP-odd states, we perform a maximum likelihood fit to the mass, lifetime, and time-dependent angular distributions of the $B_s^0 \to J/\psi (\to \mu^+ \mu^-) \; \phi (K^+ K^-)$ decay. The fit yields the CP-violating phase $\phi_s$ and the width difference $\Delta \Gamma_s \equiv \Gamma_L - \Gamma_H$. The decay can be described by three decay angles $\theta$, $\phi$, and $\psi$ defined in \cite{d0_phis_prl}. We employ initial state flavor tagging which improves the sensitivity to the CP-violating phase and removes a sign ambiguity on $\phi_s$ for a given $\Delta \Gamma_s$ present in our previous analysis~\cite{d0_phis_prl_previous}.  
In the fit, $\Delta M_s$ is constrained to its measured value (from CDF) and the strong phases are constrained to values measured for $B_d$ at the $B$\textbf{}-factories, allowing some degree of violation of SU(3) symmetry. The $B_s$ flavor at production is determined using a combined opposite-side plus same­side tagging algorithm. Confidence level contours in the $\phi_s - \Delta \Gamma_s$ plane are shown in Fig.~\ref{fig:dg_phis}(a).
\begin{figure}
\begin{center}
\begin{tabular}{c c}
	\includegraphics[width=0.5\textwidth]{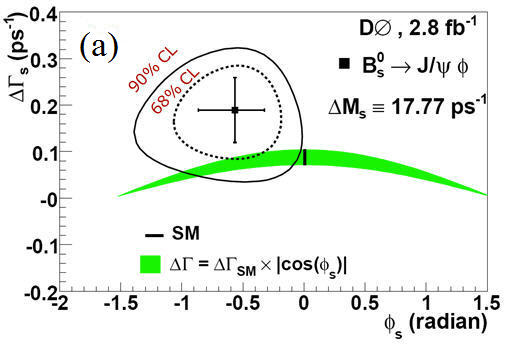} &
	\includegraphics[width=0.47\textwidth]{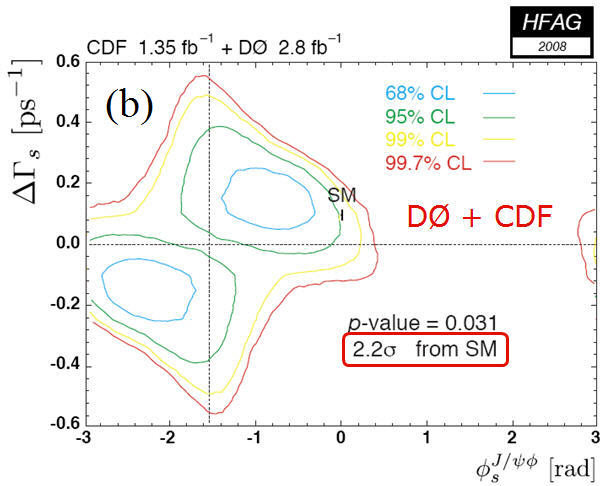}
\end{tabular}
\end{center}
	\caption{\label{fig:dg_phis}
	(a) Results of the fits in the $\phi_s - \Delta \Gamma_s$ plane for the \d0\ $B_s \to J/\psi \phi$ analysis based 
	on 2.8~fb$^{-1}$ of data. (b) Fit results from the combination of the \d0\ results with CDF results based on 
	1.35~fb$^{-1}$ of data.}
\end{figure}

The fit yields a likelihood maximum at $\phi_s = -0.57^{+0.24}_{-0.30}$ and
$\Delta \Gamma_s = 0.19 \pm 0.07$~ps$^{-1}$, where the errors are statistical only.
As a result of the constraints on the strong phases, the second maximum, at  
$\phi_s = 2.92^{+0.30}_{-0.24}$, $\Delta \Gamma_s = -0.19 \pm 0.07$~ps$^{-1}$, is disfavored by a likelihood ratio of 1:29. 

From the fit results and studies of the systematic errors we obtain the width difference 
$\Delta \Gamma_s \equiv \Gamma_L - \Gamma_H = 0.19 \pm 0.07 {\rm (stat)}\thinspace ^{+0.02}_{-0.01} {\rm (syst)}$ ps$^{-1}$ and the CP-violating phase, $\phi_{s} = -0.57 ^{+0.24}_{-0.30} {\rm (stat)}\thinspace ^{+0.07}_{-0.02} {\rm (syst)}$.
The allowed 90\% C.L. intervals of $\Delta \Gamma_s$ and $\phi_s$ are $0.06 <\Delta \Gamma_s <0.30$ ps$^{-1}$ and $-1.20 <\phi_s < 0.06$, respectively. The probability to obtain a fitted value of $\phi_{s}$ lower than -0.57 given SM contributions only is 6.6\%, which corresponds to approximately 1.8~$\sigma$ from the SM prediction.

The results have been combined with CDF results by the Heavy Flavor Averaging Group. The CDF results are based on a data set of 1.35~fb$^{-1}$ and the combination was performed with no constraints on the strong phases. The results are shown in Fig.~\ref{fig:dg_phis}(b). The fit yields two solutions as follows:
\begin{eqnarray*}
  \phi_{s} &=& -2.37 ^{+0.38}_{-0.27}~\mathrm{rad}, ~~ \Delta \Gamma_s = -0.15 ^{+0.066}_{-0.059}~\mathrm{ps}^{-1} \\
  \phi_{s} &=& -0.75 ^{+0.27}_{-0.38}~\mathrm{rad}, ~~ \Delta \Gamma_s = 0.15 ^{+0.059}_{-0.066}~\mathrm{ps}^{-1} \\
\end{eqnarray*}
The $p$-value assuming only SM contributions is 3.1\%, corresponding to 2.2~$\sigma$ from the SM prediction.

\section{Search for Direct CP Violation in {\boldmath$B^\pm \to J/\psi \; K^\pm (\pi^\pm)$} Decays}

\d0 has performed a search for direct CP violation by measurement of the charge asymmetry in $B^\pm \to J/\psi \; K^\pm (\pi^\pm)$ decays~\cite{d0_directCP}. The charge asymmetry is defined by

\begin{eqnarray*}
  A_{CP} (B^+   \to J/\psi \; K^+  (\pi^+  )) = 
  \frac{{N(B^-   \to J/\psi \;K^-  (\pi^-  )) - N(B^+   \to J/\psi \; K^+  (\pi^+  ))}}
  {{N(B^-   \to J/\psi \; K^-  (\pi^-  )) + N(B^+   \to J/\psi \; K^+  (\pi^+  ))}}
\end{eqnarray*}

\noindent
Direct CP violation in these decays leads to a non-zero charge asymmetry. In the SM there is a small level of CP violation:
$A_{CP} (B^+   \to J/\psi \; K^+) \approx 0.003$~\cite{Hou} and
$A_{CP} (B^+   \to J/\psi \; \pi^+) \approx 0.01$~\cite{Dunietz}. New physics may significantly enhance $A_{CP}$. 

After selection of $B^\pm \to J/\psi \; K^\pm (\pi^\pm)$ candidates, the invariant mass of the $J/\psi K$ is constructed
and fit to the sum of contributions from  $B^\pm \to J/\psi \; K^\pm$, $B^\pm \to J/\psi \; \pi^\pm$, $B^\pm \to J/\psi \; K^*$, and combinatorial background. From the fit we find approximately 40,000 $B^\pm \to J/\psi \; K^\pm$ candidates and about 1,600 $B^\pm \to J/\psi \; \pi^\pm$ candidates. 
To extract the charge asymmetry the sample is divided into 8 subsamples according to the solenoid polarity, the sign of the pseudorapidity of the $J/\psi K$ system, and the charge of the $K$ candidate. A $\chi^2$ fit to the number of events in each subsample yields the integrated raw charge asymmetry. This is then corrected for the asymmetry of the kaon interaction rate on nucleons to obtain the final results:
\begin{eqnarray*}
  A_{CP} (B^+   \to J/\psi \; K^ +  ) &=&  + 0.0075 \pm 0.0061{\rm{ (stat}}{\rm{.) }} \pm 0.0027{\rm{ (syst}}{\rm{.)}} \\
  A_{CP} (B^+   \to J/\psi \; \pi ^ +  ) &=&  - 0.09 \pm 0.08{\rm{ (stat}}{\rm{.) }} \pm 0.03{\rm{ (syst}}{\rm{.)}} \\ 
\end{eqnarray*} 
The results are consistent with the world average results~\cite{pdg}. The precision for $A_{CP} (B^ +   \to J/\psi \;\pi ^ +  )$ is comparable with the current world average, while for $A_{CP} (B^+   \to J/\psi \;K^ +  )$ the precision is a significant (factor of 2.5) improvement over the current world average.

\section{Search for Direct CP Violation in Semileptonic $B_s$ Decays}

In this section we report a new search for CP violation in the decay 
$B_s^0  \to D_s^-  \mu^+  \nu X, \;\; (D_s^- \to \phi \pi^- , \;\; \phi \to K^+ K^-)$
by measurement of the charge asymmetry using a time-dependent analysis with flavor tagging. The technique used is
similar to that used in the \d0\ $B_s$ oscillation analysis (see Section~\ref{sec:bosc}). 
A fit to the invariant $KK\pi$ mass of the selected data is shown in Fig.~\ref{fig:m_KKpi}.
\begin{figure}[htp]
	\begin{center}
	\includegraphics[width=0.5\textwidth]{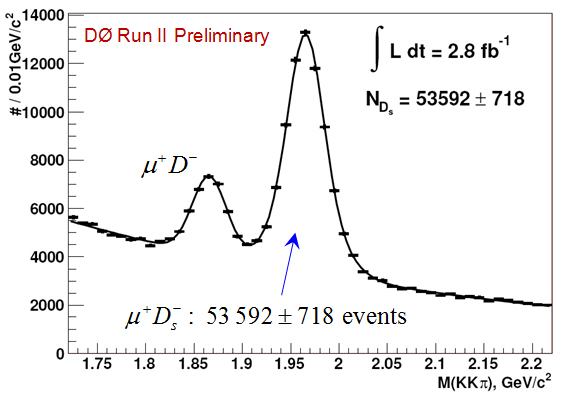}
	\caption{\label{fig:m_KKpi}
	The $KK \pi$ invariant mass distribution showing the $\mu^+ D^-$ and $\mu^+ D_s^-$ signals together with the fit results.}
	\end{center}
\end{figure}

The sample is divided into 8 subsamples according to solenoid polarity, the sign of the pseudorapidity of the $D_s \mu$ system, and the muon charge. An unbinned likelihood fit is used to extract the asymmetry. The systematic uncertainties are mainly due to uncertainties in the $c \bar c$ contribution, uncertainties in the efficiency vs visible proper decay length, and uncertainties in the $B_s \to D_s^{(*)} \mu \nu$ branching fractions. Accounting for these yields the final result: 
\begin{eqnarray*}
	a_{sl}^s  =  - 0.0024 \pm 0.0117{\rm{ (stat}}{\rm{.) }}_{ - 0.0024}^{ + 0.0015} {\rm{ (syst}}{\rm{.)}}
\end{eqnarray*}

This result is consistent with the SM prediction and is the most precise measurement to date.

\section{Measurement of $Br(B_s^0 \to D_s^{(*)} D_s^{(*)})$}

For the $B_s^0$ system, the width difference between $\Delta \Gamma_s \equiv \Gamma_L - \Gamma_H$
is related to the width difference between the CP eigenstates, $\Delta \Gamma_s^{CP} \equiv \Gamma_s^{even} - \Gamma_s^{odd}$, by $\Delta \Gamma_s = \Delta \Gamma_s^{CP} \cos{\phi_s}$. The quantity $\Delta \Gamma_s^{CP}$ can be estimated from the branching fraction $Br(B_s^0 \to D_s^{(*)} D_s^{(*)})$.


\d0\ has performed a preliminary measurement of $Br(B_s^0 \to D_s^{(*)} D_s^{(*)})$ based on 2.8~fb$^{-1}$ of data using events which contain decays $D_s^+ \to \phi \pi^+$ and $D_s^- \to \phi \mu^- \nu$ (and their charge conjugates). The signal is extracted from a 2-dimensional likelihood fit to the data in the 
$m(D_s \phi \pi) - m(\phi \mu)$ plane. The results are~\cite{Ds_prl}:
\begin{eqnarray*}
Br(B_s^0 \to D_s^{(*)} D_s^{(*)}) = 0.035 \pm 0.010 {\rm{(stat.)}} \pm 0.011 {\rm{(syst.)}} \\
\frac{\Delta \Gamma_s^{CP}}{\Gamma_s} = 0.072 \pm 0.021 {\rm{(stat.)}} \pm 0.022 {\rm{(syst.)}} 
\end{eqnarray*}

\noindent
Assuming $CP$ violation in the SM is small, the results are in good agreement with the SM prediction.

%
%
%

%
\end{document}
